\documentclass[10pt,twocolumn,letterpaper]{article}

\usepackage{graphicx}
\usepackage{subfigure}
\usepackage{psfrag}
\usepackage{textcomp}
\usepackage{amsthm}

\newcommand{\myfigure}[3]
{\begin{figure}[hbt]\begin{center}#2\caption{#3}\label{#1}\end{center}\end{figure}}

\newcommand{\mygraphfigure}[4]{\myfigure{#1}{\includegraphics[#2]{#3}}{#4}}

\newtheorem{theorem}{Theorem} 

\newcommand{\diff}[1]{\mathrm{d} #1}
\newcommand{\der}[2]{\frac{\diff{#1}}{\diff{#2}}}

\begin{document}  
\psfragscanon


\title{Exponential Tethers for Accelerated Space Elevator
Deployment\footnote{\copyright 
2004 Institute for Scientific Research, Inc.}
\footnote{This paper was first published in \textit{Proc. of the 3rd International 
Space Elevator Conference, June 2004}, reprinted on arXiv by permission of Bradley Edwards 
former director at ISR.}
}
\author{Blaise Gassend\footnote{The author may be contacted at gassend@alum.mit.edu.}}
\date{}
\maketitle

\begin{abstract}

An exponential space elevator is a space elevator with a
tether cross-section that varies exponentially with altitude. With such an
elevator it is possible to reel in tether material at one end of the
elevator while reeling out at the other end, without changing the overall
taper profile. I show how to use this property to build up or clone a space elevator
much more efficiently than with standard climber-based methods.

\end{abstract}

\section*{Introduction}

Space elevators are a promising candidate for replacing rockets as the
principal means of transportation into space. With them, space can be reached by
climbing a giant tether, attached to the Earth at one end, and held in
place by centrifugal force due to the Earth's rotation.
The concept was first proposed in Russian by
Artsutanov~\cite{artsutanov60,lvov67}, and later introduced in English by
Pearson~\cite{pearson75}. The idea was then mainly developed by
science-fiction authors, including Arthur C. Clarke~\cite{clarke79}, who said ``The
space elevator will be built about 50 years after everyone stops
laughing''. The laughing has largely stopped since a proposed light weight
elevator concept by Edwards~\cite{edwards2002}.

In his design, Edwards proposes to use carbon nanotube composites as the
building material for the elevator, the availability of sufficiently strong
composites currently being the main impediment to building the elevator.
In his proposal, the elevator construction begins with the 
launch to geosynchronous orbit (GEO) of a very light initial tether; light
enough for existing launch technology. Once at GEO, this initial tether is
deployed down to an \emph{anchor} 
station on the surface of the Earth. At the other end a
\emph{counterweight} located beyond GEO pulls on the tether to keep it upright
and in tension. This initial tether is light and therefore has a small
payload. It is nevertheless strong enough for a light \emph{climber}
vehicle to pull itself up the tether. The tether is ribbon shaped to allow the
climber to climb up it, and to protect it from space debris. 
As
it climbs, the climber uses a spool of tether material it is carrying to slightly
widen the existing cable. This climber is followed by many others, each one
helping to build up the cable a little more. After a couple of hundred
climbers, the space elevator is complete, ready to lift much larger
payloads.

Edwards, like his predecessors, considers that the tether has a
cross-section that depends on altitude. Indeed, to maximize the payload,
the cross-section of the tether should be chosen so that it is fully loaded
along its full length. Indeed, if some parts of the tether were not fully
loaded, material could be removed from those parts, resulting in a lighter
elevator with a greater lift capacity. We shall refer to these tethers as
uniform-stress tethers. They are skinny at the surface of the
Earth, increase in cross-section with altitude up to GEO, after which they
decrease in cross-section. They are the uncontested choice for making a space
elevator that lifts large payloads, and they tend to be regarded as the
only interesting tethers for space elevators.

My goal in this paper is to show that this idea is wrong, and that other
tether profiles are also useful for space elevators. Indeed, while
uniform stress-tethers are ideal for lifting payloads, they are not
necessarily the best tether profile to use while building up the elevator.
Indeed, the build up of a space elevator is limited by how fast one can get
mass off the ground onto the elevator. During build-up, the mass that needs
to be lifted is essentially made up of ultra-strong tether material.
This is very different from the type of mass that is to be lifted once the
elevator is in general use (satellites, probes, ...). In Edwards' proposal,
climbers are used to lift both types of mass. But while the tether is on the
climber, it is just dead weight; its strength is wasted.

In this paper, instead of lifting ribbon materials using climbers, we shall
lift ribbon material by adding material to the bottom of the elevator.
Think of a kite. If you want to lift string off the ground, you can either
have a roll of string lifted along the kite's string with a little climber
device, or you can simply let out more string; the kite will
rise to maintain tension in the string. In the latter case, as
material is lifted, it is also providing strength to help lift the material
below it. When this principle is applied to space elevators, we shall see
that mass can be lifted much faster than is possible using climbers.

Unfortunately, if the space elevator has a uniform-stress profile, then
this method will not work. Indeed, the thick part of the elevator that is
initially at GEO ends up past GEO, and the elevator no longer has
uniform stress. Moreover, it is not clear what profile should be
given to the new material being added at the base of the elevator.
Consequently, we shall use exponential tethers instead of
uniform stress tethers. With these tethers, the cross-section depends
exponentially on altitude. When the tether is translated up or down, the
cross-section of the elevator is simply multiplied by a constant factor;
the overall profile remains unchanged. Exponential taper is the
natural taper for tethers that get translated.

The idea of using an exponential tether that gets translated
upwards during buildup is not entirely new. To my knowledge it was first
proposed by Cline~\cite{cline2002}. The \emph{reel-to-reel} buildup method I
describe is essentially the method proposed by Cline. The \emph{breeder
elevator} method is also hinted at in his work. In writing this paper I hope
to expose a wider audience to these excellent ideas. I have also tried to
go beyond Cline's work. In particular, I feel that I have provided
quantitative arguments in favor of these buildup methods, I have studied
them over a wide range of elevator parameters, I have recognized that
exponential taper is the canonical taper profile to use, and I have
realized that the
breeder elevator can be used even when the tether material is too weak to
permit tethers with inverse taper. The \emph{pull-down} and \emph{redeploy
and splice} methods are also, new to my knowledge.

In the remainder of this paper, I shall first consider the basic equations that apply to
exponential tethers, and review some of their properties in
Section~\ref{sec:analysis}. Then I
shall present different ways in which exponential tethers can be used in space
elevator construction in Section~\ref{sec:usage}. Finally, I will compare
the proposed schemes with Edwards' climber-based scheme in
Section~\ref{sec:comparison}.

\section{Properties}

\label{sec:analysis}

Before looking at how exponential tethers can aid in space elevator
deployment, we look at some of the properties of exponential tethers. For
simplicity, in this paper we will consider that the tether has no elasticity 
and that it is located in the equatorial plane. All calculations will be
done assuming an Earth elevator (radius $r_e=6.38\cdot10^6~\mathrm{m}$, mass 
$M_e=5.98\cdot10^{24}~\mathrm{kg}$). 
We will explore a range of tether
materials. Most generally, tether material is characterized by its strength
to weight ratio. However, for ease of understanding, I will characterize a
tether material by its maximum tensile strength (after safety factor), and
keep the density fixed at $1300~\mathrm{kg/m^3}$. 


\subsection{Taper Profile}

An exponential tether has a cross-section with an exponential
dependence on position along the tether:

\begin{equation}
\label{eq:A}
A(r)=A_0 e^{\gamma r}
\end{equation}

\noindent In this expression $r$ is the distance to the center of the
planet, $\gamma$ is the exponential growth parameter of the tether, 
and $A_0$ is the cross-section that the
tether would have if it was extended to the center of the Earth.

The key property of exponential tethers is that their cross-section
increases uniformly by a factor $e^{\gamma d}$ when they are translated a distance
$d$. This makes them the
tether of choice for buildup methods like the ones we will introduce in
Section~\ref{sec:usage}.

The value of $\gamma$ determines how much the elevator tapers. When it is
positive, the elevator broadens with altitude, this is \emph{normal taper}. 
When it is negative, the
elevator gets narrower with altitude, this is \emph{inverse taper}. 
In the space elevator literature,
people usually consider the taper ratio, which is the ratio of the elevator
cross-section at GEO to its cross-section at the anchor.\footnote{For
uniform-stress elevators, the cross-section at GEO is also the maximum
cross-section of the elevator. For exponential elevators this is
no longer the case, so the taper ratio is not the ratio of maximum
cross-section to cross-section at the base of the elevator.} The taper
ratio is given by
\begin{equation}
\beta=e^{\gamma (r_g-r_e)}
\end{equation}
where $r_g$ is the distance from the
center of the Earth to the GEO. I shall present data in terms of taper
ratio, as it is easier to grasp than the growth parameter.

\subsection{Tension and Stress in the Tether}

For the tether to be in equilibrium it must satisfy Newton's second law all
along its length. This gives an equation for the tension in the tether:

\begin{equation}
\label{eq:T}
\der{T}{r}(r)=-\rho A(r) g(r)
\end{equation}

\noindent In this equation $T$ is the tension in the tether, and $g$ is
the gravitational field (negative when it is pointing towards the Earth). 
The gravitational field incorporates gravity and a centrifugal term due to the Earth's
rotation:

\begin{equation}
g(r) = -\frac{G M_e}{r^2} + r \Omega^2
\end{equation}

\noindent Here $\Omega$ is the angular velocity of the Earth's rotation and $G$
is the gravitational constant.
The gravitational field changes sign at GEO, at 
$r_g=(G M_p / \Omega^2)^{1/3}$. Therefore, the tension is maximum at that
altitude. Equation~(\ref{eq:T}) can now be integrated. However, the
solution includes the integral exponential function and is not very
insightful.

To completely know the tension in the tether, we need to specify some
boundary conditions. At the anchor point, we simply need $T(r_e)\ge0$ so that
the tether is in tension. At the counterweight, the tension in the tether
must be exactly the tension needed to counteract the gravitational force
(i.e., here mainly the centrifugal force) on the counterweight
$T(r_c)=M_c g(r_c)$. Because of this condition, the counterweight must be
beyond GEO. This makes sense, the elevator has to reach past GEO
for the centrifugal force on the elevator to counterbalance the Earth's
gravitational field. These boundary conditions will be incompatible if the
counterweight's mass is insufficient.

The stress $\sigma$ in the tether can be expressed in terms of the tension
by $\sigma=T / A$. We get an equation for $\sigma$ from~(\ref{eq:A}) 
and~(\ref{eq:T}):

\begin{equation}
\label{eq:sigma}
\sigma(r) \gamma + \der{\sigma}{r}(r)=-\rho g(r)
\end{equation}

This equation shows that exponential tethers do not usually have 
their maximum stress at the synchronous altitude. The maximum stress is
reached at the anchor, at the counterweight, or at a point where
$\der{\sigma}{r}(r)=0$. In the latter case, the maximum stress $\sigma_m$ and
the altitude $r_m$ at which it is reached are related by 

\begin{equation}
\label{eq:max-stress}
\sigma_m \gamma=-\rho g(r_m)
\end{equation}
Taking the derivative of~(\ref{eq:sigma}) at a point
where~(\ref{eq:max-stress}) holds, we find that there are no local minima
of the stress. Therefore, the maximum stress is reached at a single point
along the tether. Moreover, we note that as $\sigma_m$ increases, 
$r_m$ gets further from GEO. We are now ready for a theorem that greatly
aids in understanding the range of altitudes that can be reached by an
exponential tether. The proof of this theorem is left up to the reader.

\begin{theorem}
\label{th:max-stress}

If an exponential tether, with one extremity on each side of GEO, exceeds a
stress $\sigma_0$, then it exceeds it at one of its extremities or at
$r_0=g^{-1}(\sigma \gamma / \rho)$ (the last case only applies if $r_0$ is
within the length of the tether). 

\end{theorem}

\subsection{Critical Strength}
\label{ssec:critical-strength}

In Section~\ref{sec:usage}, a number of the applications of exponential
tethers need the tether to have inverse taper to work. Inverse
taper is only possible for sufficiently strong tethers. To find the
critical strength $\sigma_c$ beyond which inverse taper is possible, we
consider an untapered tether with $T=0$ at the anchor. In that case the
stress is maximum at GEO, from~(\ref{eq:max-stress}). We
calculate the tension at GEO by integrating~(\ref{eq:T}) with
$\gamma=0$. The stress at GEO is equal to the 
critical stress:
\begin{equation}
\sigma_c=\rho \, G M_e \left(\frac{1}{r_e}-\frac{1}{r_g}\right) + \frac{1}{2} \rho \, \Omega^2
(r_e^2 - r_g^2)
\end{equation}
For the Earth, the critical strength is $\sigma_c=63.0~\textrm{GPa}$. This is
less than the strength that is assumed by Edwards~\cite{edwards2002} for
his designs.

\subsection{Allowable Counterweight Altitudes, Taper Ratios and Tether
Strengths}

With uniform stress tethers, for a given strength tether, it is possible to
reach any altitude, and there is a unique way of doing so. Reaching
altitudes far above or below GEO will just have a high cost, measured by a large
taper ratio. With exponential tethers, the situation is much more complex, which
comes from the three places where the stress $\sigma_0$ can be exceeded in
Theorem~\ref{th:max-stress}.

Figure~\ref{fig:min-stress} shows the tether lengths that can be achieved
for an Earth space elevator with various strengths and taper ratios.\footnote{If 
elasticity were taken into account then the
elevator would be unstable when we are too near the equator or the
counterweight is too light. Taking this instability into account 
would force us to slightly increase the necessary strengths we find in this
analysis.} 
This plot was generated by considering, for a given taper ratio and
elevator length, the maximum stress that is achieved by
integrating~(\ref{eq:T}) from each end of the elevator, starting with a
tension of zero.
The bottom branch of each plot corresponds to a limit where the tension is
zero at the anchor. The top branches correspond to limits where the
tension is zero at the counterweight. For a given tether length, there is a
minimum necessary tether strength. That
minimum strength is at the intersection of the two branches.

\mygraphfigure{fig:min-stress}{width=\columnwidth}{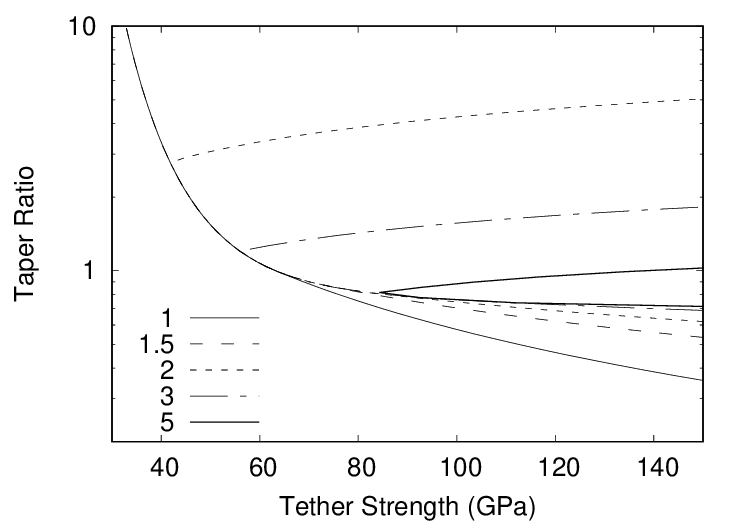}{Acceptable
combinations of counterweight position, tether strength and taper ratio.
Each line corresponds to a counterweight position (specified by $r_c/r_g$),
which can be reached in the region on the right of
the line.}

\section{Deployment Scenarios}
\label{sec:usage}

We shall now look at ways in which exponential tethers can be
used to build up a space elevator. First we shall look at
\emph{reel-to-reel} buildup in which a tether can be deployed by reeling
in material at the counterweight and letting material out at the anchor.
We will then discuss an alternative method, \emph{the pull-down} build up, 
where both reels are on the
ground and the counterweight is a simple pulley. Next we will see how 
the material that is accumulated at the counterweight could be used to build a second
elevator. This will lead to the idea of a \emph{breeder} which can be used
to rapidly produce two similarly sized elevators from one. Finally, by
merging the two resulting elevators one elevator can be caused to nearly
double in cross-section in a single step, leading to vastly improved tether
growth rates; this is the \emph{redeploy and splice} buildup method.

\subsection{Reel-to-Reel Buildup}

Reel-to-reel buildup is the most natural way to use an exponential
tether. It requires a tether with inverse taper, and thus the tether
strength must exceed the critical strength (see
Section~\ref{ssec:critical-strength}). The
buildup method is as follows (see also Figure~\ref{fig:reel-to-reel}):

\begin{enumerate}

\item An initial exponential tether is deployed.

\item The initial tether is built up by reeling in material at the counterweight 
while paying out material at the anchor. Because of the inverted taper, 
the tether cross-section slowly increases until it reaches the desired 
cross-section for the final elevator. 

\item A uniform-stress tether is reeled up, and the
elevator is ready to lift payloads.

\end{enumerate}

\mygraphfigure{fig:reel-to-reel}{width=\columnwidth}{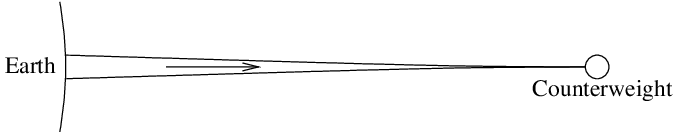}{A reel-to-reel
elevator being built up.}

\subsubsection{Advantages}

\paragraph{Speed}

For sufficiently strong tethers, (i.e., when the inverted taper 
becomes significant) this method is much faster than climber based buildup 
(see Section~\ref{sec:comparison}). 

\paragraph{Simplicity}

With this method, the only moving part is the spool at
the counterweight. This is a great simplification over having to send
hundreds of climbers up the tether.

\paragraph{Quality of ribbon}

There is no high altitude ribbon splicing, unlike
climber based buildup. This may allow a lower safety factor for
reel-to-reel buildup.

\paragraph{Ease of repair}

Ribbon repairs can be performed on the ground by reeling the
uniform-stress section of the ribbon down, and then reeling it back up
repaired. Such repairs could be done at regular intervals and would require
a few weeks of reeling time. The repairs could be done as the tether is
reeled down, or the lowered section of the tether could be replaced by a
new tether section and the old one decommissioned or repaired at leisure.

\subsubsection{Engineering Issues}

\paragraph{No Counterweight Growth}

This method makes the tether grow in a
uniform manner. However, the counterweight does not change during buildup,
except for the addition of extra tether material. Thus the initial
counterweight has to be sized for the final size of the elevator. In
particular, the reel of the initial counterweight must be sized to
accommodate all the ribbon that is to be deployed, and the structure 
of the initial counterweight must
withstand the full counterweight tension of the final elevator. This means
extra mass for the initial counterweight. This limitation can be mitigated
by building up the elevator in stages: first build up the elevator to the
limit of the initial counterweight; then pull up a uniform-stress tether;
use it to lift a new larger counterweight; pull the
uniform-stress tether section back down; finally, restart the reel-to reel
buildup; repeat as necessary.

\paragraph{Location of Power Consumption}

With climber-based buildup, most of
the power is spent near the surface of the Earth, closest to the power
beaming stations. With reel-to-reel buildup the power is all consumed in
the reeling motor at the counterweight. Power has to be beamed farther, but
at a fixed target. Alternatively, solar panels could be used at the
counterweight, but they have to be expanded as the elevator gets broader
and the expended power increases. At the anchor point, power is being
generated from the work of the tether tension.

\paragraph{Waste of Ribbon Material}

If the tether strength is not sufficiently
large compared to the critical strength (i.e., the taper is too small)
then the amount of tether material used to build the elevator gets
multiplied by a large factor. Depending on the cost of the tether
material this can greatly increase the total cost of the project. In many
cases the extra tether material can be put to use. De facto, it is used as
material for the counterweight (getting some other material to the counterweight
would actually require extra effort and time). It can also be reused as elevator material in
various ways, some of which shall be presented later in this section.
Finally, it can be used by
other space based construction projects that need high strength
fibers~\cite{cline2002}.

\paragraph{Excessive Inverse Taper of Uniform-Stress Tether} Uniform-stress
tethers can be so narrow at the counterweight end that they cannot be
pulled up without exceeding the tether strength.
In practice
this is not a significant issue; the bottom part of the final elevator can
be uniform-stress to maximize the payload to GEO while the top part is
exponential.
This hybrid configuration would only make a difference for
interplanetary payloads that benefit from riding the elevator as high as
possible.

\paragraph{Reliability of Counterweight} In reel-to-reel buildup, the
reeling mechanism at the counterweight has to function flawlessly throughout 
the buildup process. This appears to be a weakness compared with hundreds of 
climbers that each need to work for only a couple of weeks, and for which recovery 
strategies have already been proposed~\cite{edwards2002}. Nevertheless,
redundancy can be built into the counterweight, for example by having an
independent reeling system that is initially detached from the ribbon, and
that attaches itself and takes over the reeling process if the primary
system fails. Alternatively, we can try to move the reeling mechanism away
from the counterweight as will be proposed in Section~\ref{sssec:pulldown}.

\paragraph{Wind, Atomic Oxygen and Space Debris} Because each section of
cable travels from the anchor to the counterweight, each section encounters
the the whole range of perils that a space elevator must endure. 
During buildup, we can no longer use the strategy of having a thick but
narrow tether within the atmosphere against wind loading, a tether
designed to withstand chemical attack in the upper atmosphere, and a wider
tether in LEO to resist damage from space debris~\cite{edwards2002}. The
final uniform-stress tether can still have these properties, of course.

During
buildup, the atomic oxygen problem is not an issue, as each length of
tether only crosses the affected altitude range for a few hours. Wind and
Space Debris, however, must be protected against at all times as they can
cause immediate damage. Here the answer lies either in changing the tether
geometry between the upper atmosphere and LEO (the tether could be
initially rolled up laterally, and progressively unroll with altitude), or
in designing the tether so that it has both a large total width and a small
wind resistance (by having a ribbon made up of widely spaced relatively
thick threads -- this would not be as easy if the ribbon was built up from
hundreds of smaller ribbons).

\paragraph{Counterweight Altitude} As the tether is reeled in, the
counterweight's mass will change, as well as the tension applied to it,
therefore the counterweight's altitude may have to change during build up.
This aspect deserves further consideration.

\subsection{Pull-Down Buildup}
\label{sssec:pulldown}

A number of problems with reel-to-reel buildup can be solved by making a
small modification to the scheme. Instead of spooling the tether 
up at the counterweight, it goes around a pulley,
and returns to a second anchor point on Earth to be spooled
up. Figure~\ref{fig:pull-down} illustrates the scheme.

\mygraphfigure{fig:pull-down}{width=\columnwidth}{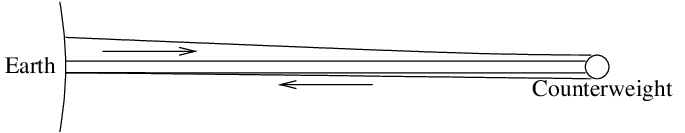}{A pull-down
elevator being built up.}

\subsubsection{Advantages}

As before, the taper cross-section increases with time because of the
inverse taper of the upwards strand of the elevator. A number of problems
with the previous design are now solved, though:

\paragraph{Location of Power Consumption} Now, the counterweight is a
passive element. All the power for getting the tether up is provided
directly by the ground station that is pulling on the down tether. In fact,
if the
counterweight is far enough beyond GEO, it is possible that the energy
needed to pull the down strand is less than the work done by tension
at the base of the up strand. In this case, build up would actually
generate energy (taken from the Earth's rotation). Whether this occurs for
realistic elevator configurations requires further study.

\paragraph{Reliability of Counterweight}
Mechanically, the counterweight is greatly simplified as it is
just a pulley. It can be designed with multiple bearings in series so that
many bearings need to fail before the pulley ceases working. 

\paragraph{Waste of Ribbon Material} Since tether
material is returned to Earth instead of being accumulated at the
counterweight, it can be immediately and easily recycled. For example ribbon
material arriving at the bottom of the down tether tether can be broadened 
and immediately sent back along the up tether.

\paragraph{No Counterweight Growth} At first sight, this is more of a
problem than before, as the counterweight is not building up mass in this
scheme. In fact, the counterweight can be made to build up mass by cutting
the ribbon in two at the counterweight. One part is reeled up by
the counterweight, and contributes to growing the counterweight; the other
goes into the down tether and allows the ground to provide the energy for the
deployment. Moreover, because the down tether has the opposite taper from
the up taper, it will be able to carry a significant payload (particularly
for large cable strengths that have a lot of taper). Thus it will be
possible to stop the elevator build up, and send a climber up the down
tether to bring necessary structural elements to the counterweight. Unlike
the simple reel-to-reel deployment, there would be no need to temporarily
deploy a uniform-stress-tether, thus saving weeks of time.

\subsubsection{Engineering Issues}

\paragraph{Getting Started}

Initial deployment is simplest if the tether is deployed with two strands,
ready to start reeling. Alternatively, a single strand can be deployed, and a
climber can be used to pull the free end of the tether from the
counterweight to the anchor, so that pull-down deployment can begin.

\paragraph{Tangled Tethers}

Because there are two tethers, there is a risk of tangling. Separating the
two anchor locations should minimize this risk, but this strategy does not
work when the tether is initially deployed. When the tethers are in motion,
the Coriolis effect can also help.

\subsection{Breeder Elevator}
\label{ssec:breeder}

One common remark that is made about space elevators is that once the first
one is built, the second one is easy. Indeed, the necessary parts for the
second elevator can be cheaply lifted into GEO by the first elevator,
without all the stringent mass limitations that apply for chemical rockets.
In fact, building the second elevator should be a top priority to avoid
having to start over from scratch if the first elevator was destroyed.
Edwards makes this point~\cite{edwards2002} and suggests that producing up
to ten cables in rapid succession may be wise.

Using climbers, a uniform-stress space elevator can clone itself in 210
days assuming the tether parameters from~\cite{edwards2002}. The simplest
way of doing this is to build up the ribbon until it has doubled in width,
and then split it down the middle (with some work needed to get half the
climbers to go with each half of the ribbon). However, as we have seen,
it is possible to get tether mass into orbit faster by reeling than by
using climbers. Therefore, we can expect space elevator cloning to be
faster by using exponential tether techniques.

Here is a proposed method for cloning an exponential elevator:

\begin{enumerate}

\item 
\label{step:pull-up}
Reel up the tether material that is needed for the new elevator
(enough to span the distance from Earth to counterweight and provide
sufficient counterweight mass for the new elevator).

\item Cut the tether at counterweight, and attach it to a new spool.
\label{step:cut}

\item Reel up a uniform-stress tether.
\label{step:up-uniform}

\item Send up the structure and machinery for the new elevator's
counterweight.
\label{step:hardware}

\item Attach the free end of the tether on the old spool to the
tether connected to Earth.

\label{step:down-uniform}

\item Reel down the uniform-stress tether.

\item Separate the two tethers at the anchor, move the new elevator to its
assigned location.

\end{enumerate}

At the end of this procedure there are two elevators of roughly
the same size. For an elevator with inverse taper, the new elevator is
larger than the initial one because in
Step~\ref{step:pull-up} the elevator is effectively being built up.
It is not necessary, however, that the initial elevator have inverse
taper. It is perfectly possible to clone an exponential taper that is
narrower at its base by this method. In this case the new elevator would
have a smaller cross-section than the old one, and could be built up
with more reeling.

This method is good because it is fast. It takes into account the fact that
tether material can be lifted faster than random hardware by using the
reel-to-reel method. Random hardware would have to be lifted on climbers.
To optimize the method, the mass of the structure and machinery for the new
elevator has to be small. As much of the new elevator's mass
as possible should be tether material. The total cloning time should be
on the order of four to five times the time to reel a full anchor to counterweight
length of ribbon, based on steps \ref{step:pull-up}, \ref{step:up-uniform},
\ref{step:hardware} and \ref{step:down-uniform}. This should take under
4 months with parameters similar to those in~\cite{edwards2002}. Most of
this time is spent getting the new counterweight in place. If a nearby
uniform-stress elevator is around it can be used to lift the new
counterweight to GEO, after which the counterweight can climb the exponential tether to
the old counterweight. Alternatively, multiple clonings can be done at
once, avoiding the need to deploy the uniform-stress tether more than once.
Because of its ability to spawn new elevators at a high rate, I have called
this elevator a \emph{breeder elevator}.

One big worry point with this method is step~\ref{step:cut}, which involves
cutting the cable at the counterweight. The robotics that carries out this
step will have to be designed very robustly to guarantee that the
counterweight does not accidentally ``let go'' of the cut tether that
connects it to Earth.

Numerous variants of the breeder elevator can be considered. The one that
is presented here was chosen mainly for its simplicity.

\subsection{Redeploy and Splice Buildup}

We have just seen that is possible to use material spooled up at the
counterweight to build a new elevator. Better yet, it is possible to use
that material to further build up an existing elevator. This leads to a
very fast build up method. As we shall see in Section~\ref{sec:comparison},
this method remains competitive even for much weaker tether materials than
those in~\cite{edwards2002}. This is in sharp contrast with climber
buildup methods that suffer significantly if the tether strength is
reduced. 

In this method, tether material that gets spooled up at the counterweight
is redeployed and spliced to the existing tether, resulting in a tether
nearly twice as wide as before. Buildup proceeds as follows:

\begin{enumerate}

\item Reel up enough tether material to reach from anchor to counterweight
and to provide additional mass to counterweight.

\item Cut the tether at counterweight, and attach it to a new spool.

\item Reel up enough tether material to reach from anchor to counterweight.

\item Attach the of the end of the tether on the old spool to the
tether connected to Earth.

\item Pull attachment point back down. The counterweight splices the
material from the two spools together as they are reeled out.

\end{enumerate}

The same remarks apply as for the breeder elevator. The main difference is
that we now have to splice tethers at the counterweight. Consumables will
probably be needed to do the splice operation. If they are light enough,
these consumables can be attached to the tether coming from Earth, and reeled up with it.

\section{Evaluation}
\label{sec:comparison}

In this section, we shall compare climber based, reel-to-reel and redeploy
and splice buildup methods. I have kept the analysis as simple as possible,
so these results are not exact. I have tried to be optimistic when evaluating
the climber based method, and pessimistic for the exponential tether
methods. I have chosen the same velocity $v=200~\mathrm{km/h}$ (climber
speed or reeling speed), as the technological limitations on the velocity
seem to be the same in all three cases. The results are strongly in favor
of the exponential tether results, despite this bias.

This study assumes that the maximum stress is the same for exponential and
uniform-stress tethers. Since exponential tethers are under-stressed over
most of their length, it may turn out that smaller safety factors can be
used for them, which would make them even more attractive.

For each method, buildup proceeds exponentially with a growth rate that we
compute. We shall not consider effects that only occur at the beginning or
end of buildup as they are negligible for large amounts of buildup. The
results are plotted in Figure~\ref{fig:comparison}. For ease of
interpretation the growth rates $\Gamma$ have been converted to doubling
times $t_2=\ln(2)/\Gamma$.

\subsection{Climber Based Buildup}

For climber based buildup, we take the ratio of mass per unit time going up
the elevator to elevator mass, to evaluate the tether growth rate. The
elevator mass is minimized when the counterweight is infinitely distant, so
we place ourselves in that case. The elevator mass is computed by
integrating the analytical expression for the
cross-section~\cite{edwards2002,pearson75}.

It has been noted~\cite{edwards2002,westling2003} that the mass rate for
climbers is better when frequent light climbers are used, rather than large
climbers that take up the whole capacity of the elevator. Let $d$ be the
distance between successive climbers. The payload capacity $\sigma A(r_e)$ 
is split between all the climbers that are below GEO. When a climber is
just departing from the anchor, the weight acting on all these climbers is
\begin{equation}
\sum_{n=0}^{\lfloor \frac{r_g - r_e}{d} \rfloor} g(Re + n d) M_{cl}
\end{equation}
The climber mass $M_{cl}$ should be chosen so that this weight equals the
payload capacity. The mass going up the elevator per unit time is simply
$M_{cl} v / d$.

Combining these results, we get a growth rate of
\begin{equation}
\Gamma_{\mathrm{cl}}=\frac{\sigma v}{d \sum_{n=0}^{\lfloor \frac{r_g - r_e}{d}
\rfloor} g(Re
+ n d) \int_{Re}^{\infty} \rho \frac{A(r)}{A(r_e)} \diff{r}}
\end{equation}

Figure~\ref{fig:comparison} shows the resulting growth rates. I have
plotted a number of different climber spacings. Edwards~\cite{edwards2002}
spaces climbers apart by three days. Closer spacings have also been shown, as
they greatly improve the growth rate. However, it isn't clear that faster rates will
be easy to achieve in practice as they involve more climbers, lighter
climbers and more frequent climber launches. The results shown here are a
bit better than in~\cite{edwards2002}, the reason for this mismatch is not
clear as Edwards does not detail how his numbers were arrived at.

\mygraphfigure{fig:comparison}{width=\columnwidth}{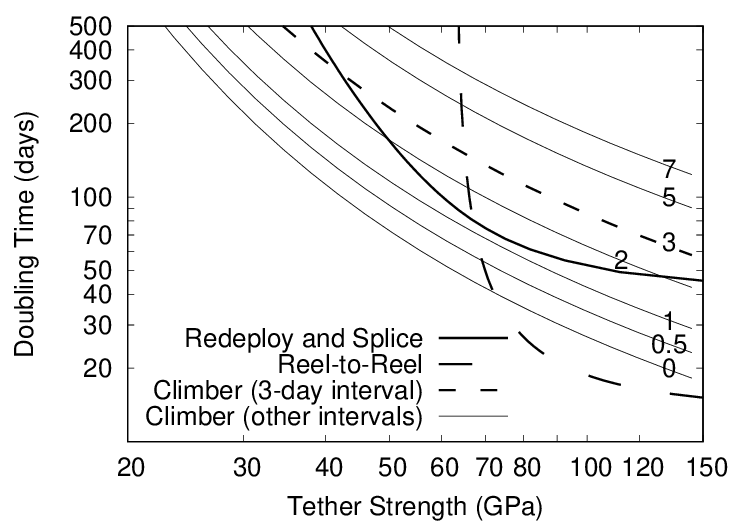}{The doubling
time for the different buildup methods, as a function of tether strength.
Assumes an Earth space elevator, and a tether density of $1300
\mathrm{kg/m^3}$. The climber method has been evaluated for various climber
departure rates indicated on the plot in days.}


\subsection{Reel-to-Reel Buildup}

For reel to reel buildup, the growth rate is simply $-\gamma v$. From
$\gamma$ we compute the minimum strength that is necessary for reel-to-reel
buildup. For simplicity, we consider that the counterweight is located in
such a way that it remains in place as the tether is reeled up. This occurs
when the mass of the counterweight is $M_c=-\rho A(r_c)/\gamma$. This is the mass
that an infinitely long extension of the tether, rolled up at the
counterweight, would have. Assuming the best case of zero tension at the 
anchor, we integrate~(\ref{eq:T})
from the anchor until the counterweight boundary condition is satisfied
with the desired mass. This condition is $-\rho A(r_c) g(r_c) = T(r_c)
\gamma$. For taper ratios less than 0.691 this condition is never satisfied, so there
is a lower bound to the amount of inverse taper we can make use of. The
value of this bound is planet dependent.

Figure~\ref{fig:comparison} shows the results. Reel-to-Reel buildup is the
absolute best choice for strengths greater than 72~GPa. At 65~GPa it is
nearly identical to climber-based buildup with a three day climber interval.
This is remarkable, given how close 65~GPa is to the critical strength of
63~GPa (see Section~\ref{ssec:critical-strength}).


\subsection{Redeploy and Splice}

For redeploy and splice, we need to determine how far out to place the
counterweight. I have chosen to place it at the point where the
counterweight mass goes to zero, to avoid having to consider the amount of
extra tether to reel for growing the counterweight. For a given taper ratio, this is
also the farthest the counterweight can be placed without needlessly
increasing the tension at the anchor beyond zero, so it is a conservative
assumption.

After one full redeploy and splice cycle, the tether has been grown by a
factor $e^{-\gamma (R_c-R_e)}+1$. The full cycle takes a time $3 (R_c -
R_e) / v$ for all the reeling operations. In practice, the counterweight
would actually move closer to the Earth during the reeling operations, so
the factor of three is pessimistic. 

The resulting growth rate is
\begin{equation}
\Gamma_{\mathrm{rs}}=\frac{v \ln(e^{-\gamma (R_c - R_e)}+1)}{3 (R_c - R_e)}
\end{equation}

Figure~\ref{fig:comparison} shows the results. Redeploy and splice is
always outperformed by the best climber based methods. However, if climbers
can only depart every three days, redeploy and splice is the method of choice
from 42~GPa to 67~GPa. I recall Dr. Edwards wondering
at the Second Space Elevator conference what could be done if the tether
strength dropped significantly below 65~GPa, as the climber method
drastically slows down past that point. The redeploy and splice method does
as well at 51~GPa as the climber method does at 65~GPa, 
so it could be the answer.

\section*{Conclusion}

In this paper I have introduced exponential space elevators, and
studied their basic properties. More importantly, I have shown how
these elevators can be used to greatly accelerate the build up phase of
space elevator construction. The simple reel-to-reel technique is well suited
to strong tethers, while the redeploy and splice method is better suited to weaker 
tethers. With the reel-to-reel technique, a $51~GPa$ tether
can be built up as fast as a $65~GPa$ tether with the currently
accepted climber based method. This could mean the difference between
feasible and impossible if carbon nanotube materials don't reach expected
strengths. At the very least, reducing the buildup time will reduce the
amount of time during which the elevator is thin and vulnerable. These new
buildup methods also lead to an elevator with a much better ribbon quality,
since the final ribbon is entirely pulled up from the ground, rather than
being spliced together by climbers at 200~km/h.

Exponential tethers have also been proposed as a faster way to produce a
new elevator from an existing one. A breeder elevator could be made that
produces a new elevator every few months. All these applications of
exponential tethers rely on the fact that mass can be lifted faster by
reeling tether material up on an exponential elevator than by using
climbers on a uniform-stress elevator. Uniform-stress elevators are the
right solution for lifting arbitrary payloads, but when ultra-strong tether
material needs to be lifted, exponential tethers are better. Exponential
elevators could have uses beyond elevator deployment, to get construction
material to other space projects.

The analysis that is presented here ignores the effects of elasticity, and
does not take dynamic effects into account. These will have to be considered
to build confidence in the new methods. More effort also needs to be put
into pull-down elevators, and the study of the counterweight motion when
exponential elevators are used in real-life scenarios.

Today the materials needed to build the space elevator are not yet
available. While we wait, we should work on lowering the minimum
strength that is needed for space elevator construction. In this paper I
have shown how getting rid of uniform-stress tethers can be a step in that
direction. I can only encourage the reader to think of other common assumptions that
can be knocked down to bring the space elevator closer to reality.

\section*{Acknowledgements}

I would like to thank all the members of the space-elevator Yahoo group for
interesting discussions, and for bringing Cline's work~\cite{cline2002} to
my attention. In particular, I would like to thank Gary Mulder, Greg
Broomfield and Val\'erie Leblanc for pointing out bugs and providing 
insightful comments.

\bibliographystyle{IEEEtran}
\bibliography{paper}

\end{document}